\documentclass[reprint,amsmath,amssymb,aps,pre,showpacs,twocolumn,floatfix]{revtex4-1}
\usepackage{epsfig}
\usepackage{graphicx}
\usepackage{makeidx}
\usepackage{chemarr}
\usepackage{dcolumn}
\usepackage{bm,multirow}
\usepackage{color}
\def\phd#1#2#3{{ Physica } {\bf #1D}, #2 (#3)}
\def\prl#1#2#3{{ Phys. Rev. Lett.} {\bf #1}, #2 (#3)}
\def\pla#1#2#3{{ Phys. Lett. } {\bf #1A}, #2 (#3)}

\def\pre#1#2#3{{ Phys. Rev. E} {\bf #1}, #2 (#3)}

\def\ptr#1#2#3{{ Phil. Trans. R. Soc. A.} {\bf #1}, #2 (#3)}
\def\ptp#1#2#3{{ Prog. Theo. Phys.} {\bf #1}, #2 (#3)}

\def\w{\omega}

\def\ep{\epsilon}

\def\etl{$et ~al.$}

\def\ep{\epsilon}

\def\bc{\begin{center}}
\def\ec{\end{center}}
\def\eqn{\end{equation}\noindent}
\def\eqnr{\end{eqnarray}\noindent}
\def\beqr{\begin{eqnarray}}

\begin{document}
\title{Amplitude death phenomena in delay--coupled Hamiltonian systems}
\author{Garima Saxena$^1$, Awadhesh Prasad$^{1}$ and Ram Ramaswamy$^{2}$}
\affiliation{$^1$Department  of Physics  and Astrophysics,
University of Delhi, Delhi 110007, India,\\
$^2$University of Hyderabad, Hyderabad 500 046, India. }

\begin{abstract}
Hamiltonian systems, when coupled {\it via} time--delayed interactions, do
not remain
conservative. In the uncoupled system, the motion can typically be
periodic, quasiperiodic or chaotic.
This changes drastically when delay coupling is introduced since now
attractors can be created in the
phase space. In particular for sufficiently strong coupling there can be
amplitude
death (AD), namely the stabilization of point attractors and the cessation
of oscillatory motion.
The approach to the state of AD or oscillation death is also accompanied by
a phase--flip in
the transient dynamics. A discussion and analysis of the phenomenology is
made through
an application to the specific cases of harmonic as well as anharmonic
coupled oscillators, in particular the H\'enon-Heiles system.\end{abstract}
\maketitle

\section{Introduction}

The coupling between dynamical systems can give rise to a number of
collective phenomena such as synchronization, phase locking,
phase shifting \cite{sync}, amplitude death \cite{yam,eli,ins,resmi,rep}, 
phase-flip \cite{pf,pfb,nirmal}, hysteresis \cite{hys}, riddling \cite{rid} and so on \cite{gen}. 
Since communication between the individual
systems is mediated by signals that can have a finite transmission time,
many studies account for this by the introduction of time--delay in the
coupling \cite{wagner,konishi,reddyall,zou,ravi,senthil}. 
A number of recent studies have examined the manner in which
delay coupling can affect the collective dynamics, particularly since
time--delay makes the systems effectively infinite dimensional \cite{farmer}.

When conservative systems are coupled via time--delayed interactions, then
there are additional considerations. To start with,  the system becomes 
explicitly non--conservative and thus the nature of the dynamics changes drastically: 
in the uncoupled system, the phase flow preserves volumes \cite{tabor}, but in the 
coupled system there can be attractors. This issue is of  added interest when studying 
Hamiltonian systems where there can be a hierarchy of conserved quantities \cite{arnold}. 
Studies of coupled Hamiltonian systems have largely examined the case of instantaneous 
coupling \cite{zanette} which does not affect the Hamiltonian structure.

In the present work we study time--delay coupled Hamiltonian systems and examine the effect of interaction on the  
nature of the dynamics. We consider the following examples,
that of diffusively coupled harmonic oscillators that models delay--coupled pendulums, for instance, 
and coupled H\'enon--Heiles oscillators. In the absence of coupling, in
the former case the motion is periodic, while in the latter case the
dynamics can be (quasi)periodic or chaotic. In both instances we find
that the effect of introducing dissipation is to cause the oscillatory
dynamics to be damped to a fixed point, namely we find that there is the
so--called amplitude death (AD) \cite{rep} as has been seen in delay--coupled 
dissipative dynamical systems \cite{rep,reddyall}.  

Although the major effect of the coupling is to make the overall system dissipative,
there are differences from the case when non-Hamiltonian systems are coupled. 
When the dynamics is decaying to a point attractor, there is an abrupt transition 
in the relative phases of the oscillatory transient motion. This is the phase--flip transition
that has been seen in a number of other systems \cite{nirmal}. Here, however, 
there are special values of the time--delay when the coupling term
effectively vanishes: the underlying Hamiltonian structure then becomes apparent.

Our main results are presented in Sections II and III where we consider
the cases of coupled harmonic oscillators and coupled H\'enon
Heiles systems respectively. We show how AD is reached and the nature of the
phase-flip transition in both cases.  Since the uncoupled systems are Hamiltonian,
it is possible to define an energy, and while this quantity has been studied in coupled feedback 
oscillators \cite{wang} as a tool to determine onset of AD, its variation in the AD regime itself has not been
examined. In this work we do energy analysis in the AD region and find that the energy dissipation is non monotonic as a function of the coupling,
decaying faster prior to the phase--flip transition and slower subsequently. The paper concludes
in Section IV with a summary and discussion of the results.

\section{Delay coupled harmonic oscillators}

The simplest system we consider is that of  diffusively coupled harmonic
 oscillators. We consider the following equations of motion
\begin{eqnarray}
\ddot{x}_j+\omega_j^2 x_j-\ep [\dot{x}_k(t-\tau)-\dot{x}_j(t)]=0
\label{eq:shm}
\end{eqnarray} \noindent
where $j, k$=1, 2 and $j \ne k$  and $x_j$ and $\dot{x}_j$ represent the
position and the velocity of the $j$th oscillator, and $\omega_j$ is the
intrinsic frequency. We take the oscillators to be identical \cite{mis},
 $\omega_1=\omega_2=\omega=1$. The parameters $\ep$ and $\tau$
represent coupling strength and time delay respectively.  
In the absence of delay, $\epsilon$ causes the systems to synchronize completely. Due to the simple dynamics 
of the system no other significant behavior is observed. When, delay is finite the coupling quenches oscillations leading to AD. 

\begin{figure}
\includegraphics [scale=0.5]{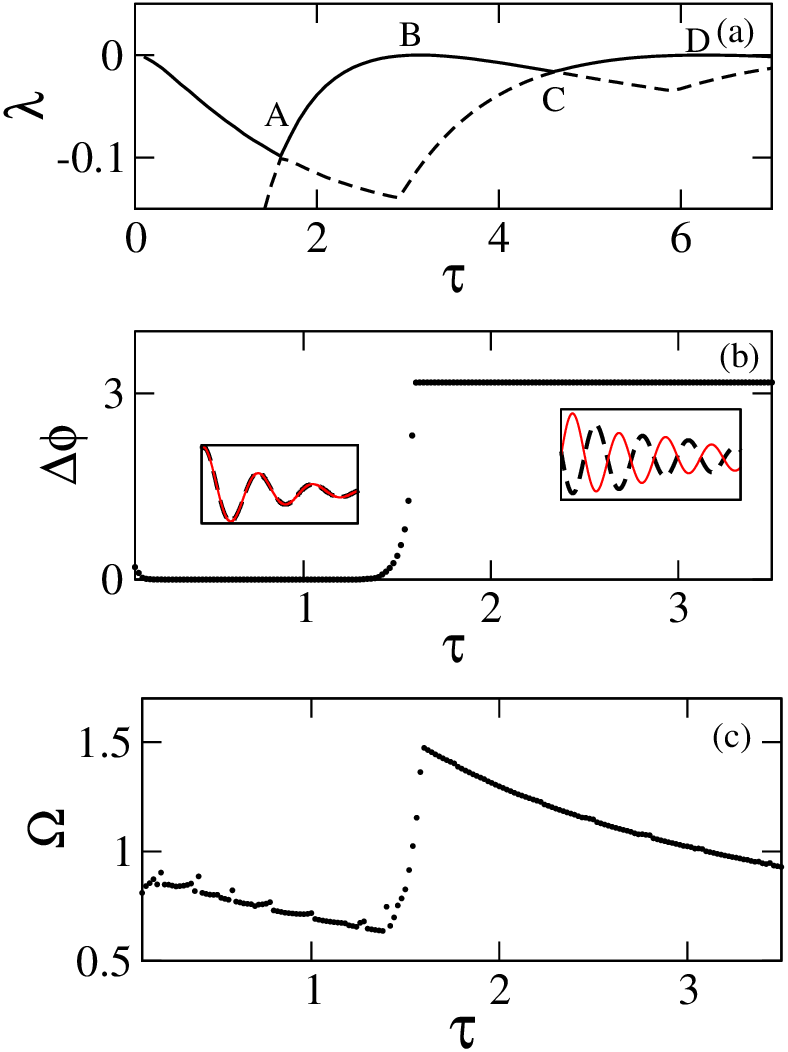}
\caption{ (Color online) (a) Lyapunov exponents  of
 coupled harmonic  oscillators Eq. (\ref{eq:shm}) as a function of the
time--delay $\tau$ for fixed $\epsilon$ = 1;
 (b) phase difference between the oscillators around point A indicated in
(a). Inset figures show  trajectories $x_1$ and $x_2$ as a function of time
for the
in--phase and the out--of--phase dynamics at $\tau$ =1 and 2 respectively,
namely on either side of the phase--flip transition,  and (c) the frequency
of oscillation of the subsystems.}
\label{fig:ho1}
\end{figure}

Stability analysis of Eq.~(\ref{eq:shm}) around the  fixed point, namely
the origin,  gives the characteristic equations
\begin{eqnarray}
\lambda (\lambda+\ep)+\omega^2=\pm \ep \lambda \exp{(-\lambda\tau)}.
\label{eq:ch}
\end{eqnarray}\noindent
Taking the roots of the Jacobian to be $\lambda=\alpha+ i \beta$, the
condition for
marginal stability is $\alpha=0$, and substituting this condition in
Eq.~(\ref{eq:ch}) we get
\begin{eqnarray}
\tau &=&\tau_c=\frac{n\pi}{\w}=\frac{nT}{2} \mbox{~~and}\nonumber \\
\beta &=& \pm \w \nonumber
\label{eq:ch1}
\end{eqnarray} \noindent
where $\tau_c$ is the critical value at which $\alpha=0$ and $T$ is the
time-period of the uncoupled oscillators. Shown  in Fig.~\ref{fig:ho1}(a)
are the first two Lyapunov exponents \cite{num} of the system as a function of $\tau$
for fixed coupling strength $\ep$ = 1. The largest LE  is zero only at
$\tau_{c} = nT/2$ (marked in Fig.~1(a) as B and D) and remains negative for all other
values of $\tau$, implying that the system is driven to  AD except when the
delay is an integral multiple of half the time period.
Further, at these critical delay values the system oscillates at the
frequency of the uncoupled  system namely $\beta = \w$, and the parameter
space is divided into multiple AD regions by the critical delay values
which are independent of  the coupling strength $\ep$. In contrast,  
in non--Hamiltonian systems AD islands are separated by finite range of delay
 values \cite{rep,reddyall}  where the coupling function need not vanish.  
Hence, in those systems the reappearance of oscillations after AD depends both on the  
coupling strength and on the delay, whereas  in coupled Hamiltonian systems
 we find that this happens only due to delay. 

In the AD regime(s)  points of discontinuity in the slope of the largest LE
 (marked in Fig.~1(a) as A and C) indicate the change in the relative
phases of oscillation. This is the so--called phase-flip transition
 \cite{pf}, and the difference in the phases of the coupled
subsystems changes by $\pi$; see (Fig.\ref{fig:ho1}{b}). As in other cases
where this phenomenon has been observed, there is simultaneously a
discontinuous change in the oscillation frequency \cite{fqcal}, as shown in
Fig.\ref{fig:ho1}(c).

\begin{figure}
\includegraphics [scale=0.4]{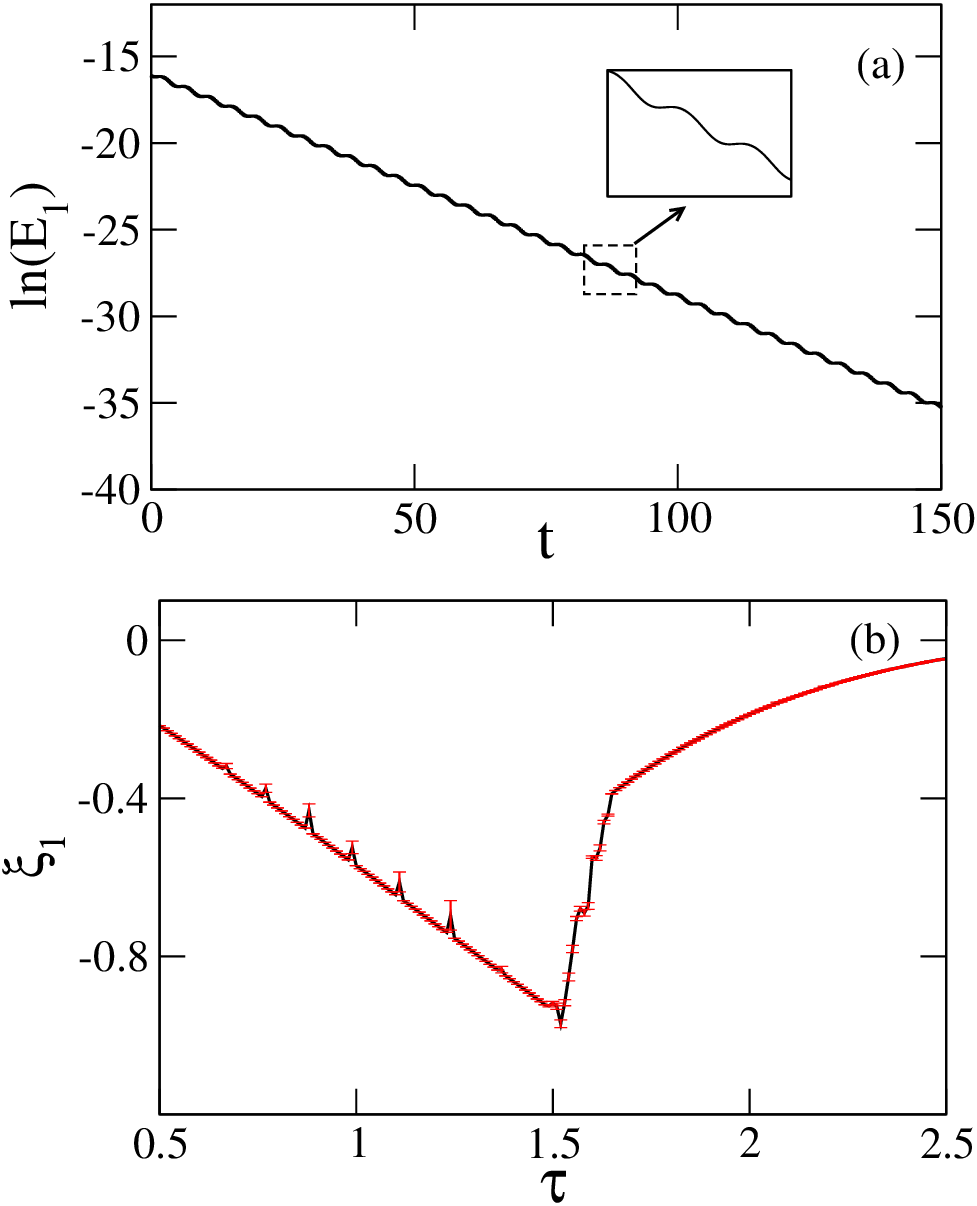}
\caption{ (Color online) (a)Energy of one of the oscillators, Eq. (\ref{eq:en}), as a
function of time for $\tau=1$.  The marked  box is expanded in the inset.
 (b) The decay rate, $\xi_1$ as a function of $\tau$ (see  Eq.
(\ref{eq:decay})). Errorbars (marked in red)  are calculated for 100 initial 
conditions.}
\label{fig:ho2}
\end{figure}

The uncoupled conservative Hamiltonian systems are made dissipative through
the time--delay coupling, and this is also reflected in the fact that the
sum of all the LEs remains  negative for
all $\tau$. Defining the energy of the individual oscillators as
\begin{eqnarray}
E_j=(\dot{x}_j^2+\Omega_j^2 x_j^2)/2
\label{eq:en}
\end{eqnarray}
where $\Omega_j$ is the instantaneous frequency of oscillation, the
approach to the fixed point can be seen to be at an exponential rate in the
AD regime \cite{trans} as can be seen in Fig. \ref{fig:ho2}(a). The
exponential decay is however modulated, the oscillatory behavior being due
to coupling (see the inset in the figure). In order to capture the
dynamics, we define a decay constant as
\begin{align}
e_j &= \langle\,\ln|E^{m+1}_j-E^{m}_j|\rangle_{m} \nonumber \\
\xi_j &= \langle\,e_{j}\,\rangle_{IC}
\label{eq:decay}
\end{align}
where  $E^m_j$ represents the $m$-th maxima in the energy time series of
the $j$th oscillator.
The averages $\langle\cdot\rangle_m$ and $\langle\cdot\rangle_{IC}$ are
performed on
$m$ and 100 initial conditions respectively.

Since the oscillatory behavior is modulated by exponential decay,
$E^{m+1}_j-E^{m}_j$ is
also an exponentially decaying function of $m$, at rate $\xi_j$.  This rate
can be measured experimentally and its variation with $\tau$ is shown in
Fig. \ref{fig:ho2}(b); the variation
mirrors that of the frequency change at phase--flip, suggesting that energy
decays more rapidly before the transition than after it.

\begin{figure}
\includegraphics [scale=0.4]{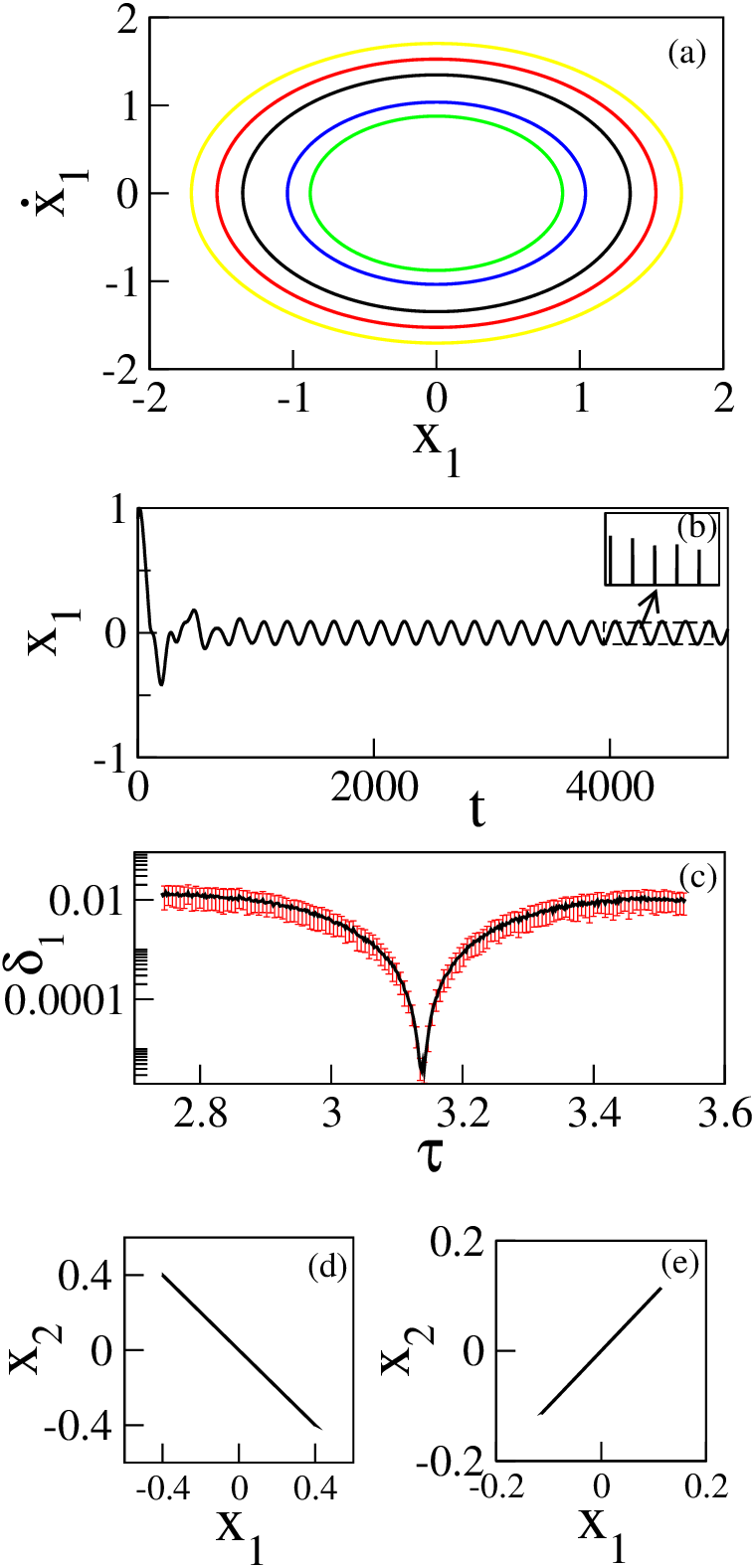}
\caption{ (Color online) (a) Periodic orbits with five different
initial conditions at $\tau=\pi/\omega$;
(b) time series of position variable $x_1$ as a function of time at $\tau=3.14$. The inset
figure shows decrease in amplitude of $x_1$;
(c) The averaged distance $\delta_1$ with errorbars, as function  of delay
$\tau$ near $\tau=\pi/\omega$.
The errorbars are calculated for 100 initial conditions;
(d) out-of-phase motion and (e) in-phase motion in relative phase plane
$x_1-x_2$ at $\tau=\pi/\omega$ and $2\pi/\omega$
 respectively (corresponding to points B and D of Fig. \ref{fig:ho1}(a)).}
\label{fig:ho3}
\end{figure}

At the critical delay values  $\tau=n\pi/\omega$ (see points
B and D in Fig. \ref{fig:ho1}(a)) the largest  Lyapunov exponent is zero and thus the motion
is periodic. Fig. \ref{fig:ho3}(a) shows orbits for five
different initial conditions at the critical point; these resemble invariant 
curves as in conservative systems. However, in the vicinity of the critical delay values the 
largest lyapunov exponent is near zero and the motion appears periodic after an initial transient. 
The time series of one such periodic orbit 
is shown in Fig. \ref{fig:ho3}(b). Also, as can be seen in the inset,
there is very slow decrease in amplitude. This occurs since delay is not strictly $\tau_{c}$. 

Away from delay $\tau_{c}$,
the dissipation is more pronounced. The rate of decrease of the amplitude
can be quantified through the measure

\begin{align}
d_j &=\langle\,|x_j^{m+1}-x_j^m|\,\rangle_{m} \nonumber \\
\delta_j &=\langle d_j \, \rangle_{IC}
\label{eq:dis}
\end{align}
where $x_j^m$ is the $m$-th maxima of $x_j$. This is plotted in Fig.
\ref{fig:ho3}(c) as  a function of $\tau$ in the vicinity of
 $\tau=\pi/\omega$, namely the point B in Fig. \ref{fig:ho1}(a). At
$\tau=n\pi/\omega$ the rate of decrease of amplitudes approaches zero and
hence the orbits are  almost periodic. Similar behavior is observed at
point D of Fig.\ref{fig:ho1}(a).

The reason for reappearance of oscillatory motion is straightforward. When the delay is a
multiple of half the natural period of oscillation, then the coupling term
effectively vanishes since
\begin{eqnarray}
\dot{x}_k(t-\tau) \approx \dot{x}_j(t)
\end{eqnarray}
and the system effectively becomes conservative. Clearly when this occurs,
each initial condition gives rise to an invariant curve (or, in this case,
a nearly invariant curve). The better the equality above is realized, the
more long lived the transients are.

There is the phase--flip transition at A, C and also higher values of $\tau$. At each transition a phase difference of $\pi$ is introduced between the oscillators resulting in anti--phase synchronization at $\tau=T/2$ and in--phase synchrony as $\tau=T$ and so on. Hence, the coupling function becomes zero at these critical delays.
The consecutive oscillatory states alternate between having the oscillations
being in--phase or out--of--phase, as shown in
Fig. \ref{fig:ho3}(d) (out--of--phase at $\tau=\pi/\omega$, namely at  B) and
 Fig. \ref{fig:ho3}(e) (in--phase at  $\tau=2\pi/\omega$ at the point
marked $D$) respectively.  Note that the phase-relation between oscillators
is independent of the initial conditions.

\section{Coupled H\'enon-Heiles systems}

In order to examine the dynamics when the uncoupled systems are capable of
exhibiting chaotic motion, we examine the behavior of two non-integrable
H\'enon-Heiles systems \cite{tabor},
\begin{eqnarray}
\nonumber
\ddot{x}_j+x_j+2x_jy_j-\epsilon (\dot{x}_k(\tau-t)-\dot{x}_j(t))&=&0
\nonumber \\
\ddot{y}_{j}+y_j+x_j^2-y_j^2&=&0
\label{eq:hh}
\end{eqnarray}
\noindent

\begin{figure}
\includegraphics [scale=0.45]{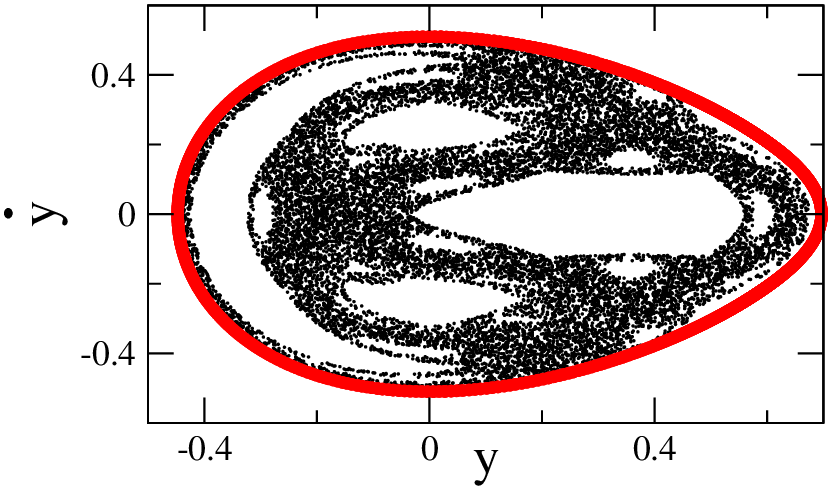}
\caption{ (Color online) (a) Poincar\'e map of two orbits in the
H\'enon-Heiles system, Eq. (\ref{eq:hh}),
at energy $E_j=0.13$. Red (outer) and black (inner) dots represent regular and chaotic
motion respectively.}
\label{fig:hh0}
\end{figure}

As is well--known, in the uncoupled case ($\epsilon$=0) the system has both
regular and irregular
behavior largely depending on the total energy as well as on the  initial
condition \cite{tabor}.
Shown in Fig.~\ref{fig:hh0} are  the Poincar\'e maps for two different
initial conditions,
one leading to regular motion (red points), while another leads to chaotic
dynamics (black points),
at the same energy $E_j$ = 0.13 just below the dissociation limit $E_j$ =
1/6.

\subsection{Instantaneous coupling $(\tau=0)$}

\begin{figure}
\includegraphics [scale=0.3]{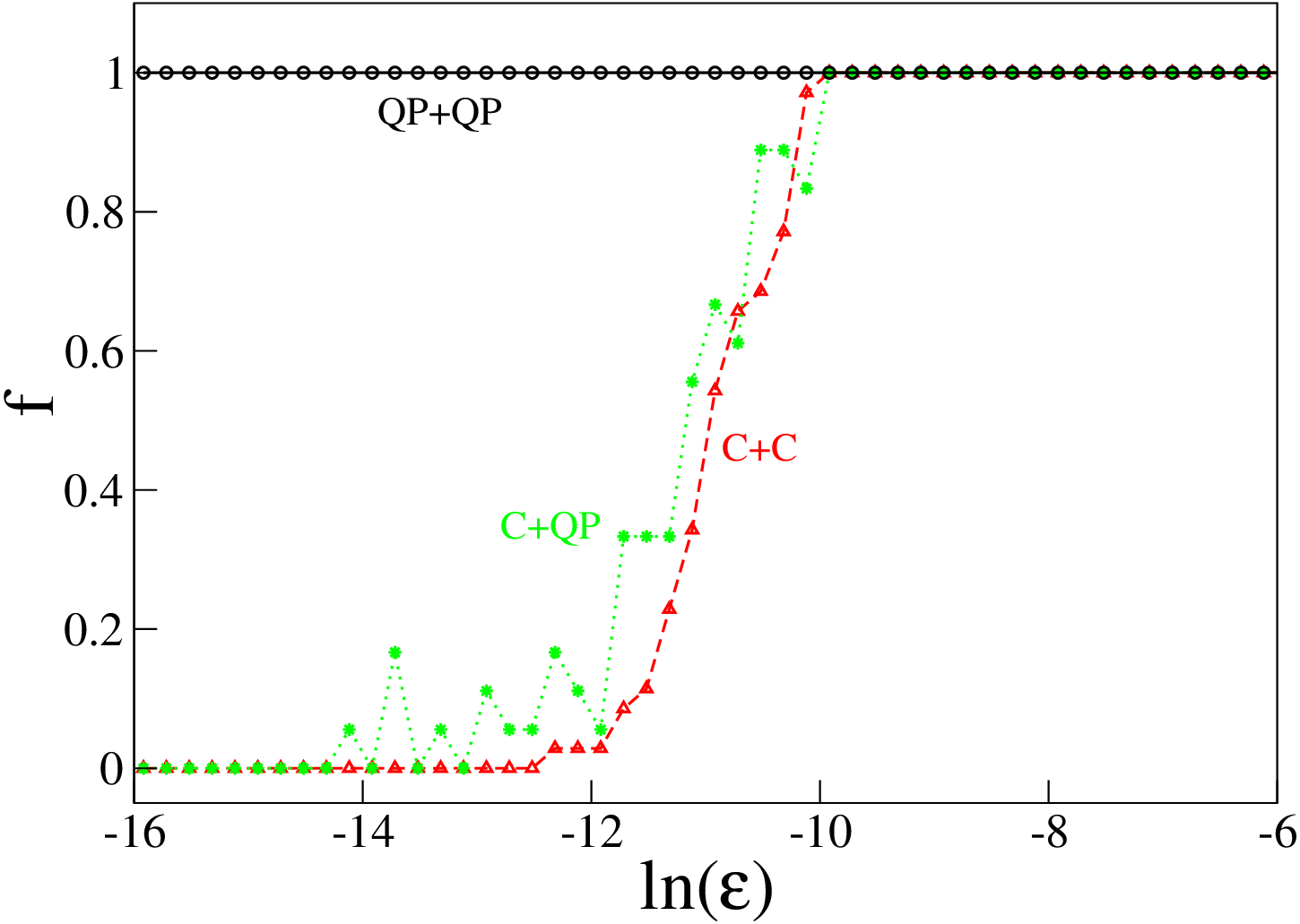}
\caption{(Color online) Fraction of initial conditions that lead to quasiperiodic  motion. 
The solid (black circles), dashed (red triangles) and dotted (green stars) curves are the results for 
the different cases of initial conditions corresponding to quasiperiodic ($QP+QP$), chaotic ($C+C$) and 
mixed chaotic and quasiperiodic $(C+QP)$ motions respectively. Averages have been taken over a sample
of 100 initial conditions.}
\vskip 0.2cm
\label{fig:hhic}
\end{figure}

In the absence of delay $\tau=0$, Eq. (\ref{eq:hh}) reduces to the case of simple diffusive coupling. 
The effect of increasing the coupling strength, namely $\epsilon$, is to induce simplicity to the resulting collective dynamics. 
Shown in  Fig.(\ref{fig:hhic}) are the the fraction of initial conditions $(f)$ leading to quasiperiodic motion.
We take 100 pairs of random initial conditions from the bounded region of phase space (Fig. \ref{fig:hh0}). 
In one case, when the initial condition pairs of quasiperiodic motions are taken, the collective dynamics 
due to interaction remains quasiperiodic (solid line (black circles): $QP+QP$). However, if  the initial motion is chaotic 
then the resulting dynamics becomes quasiperiodic only after certain  value of coupling strength (dashed line (red triangles): $C+C$). 
Similar behavior is observed when initial conditions of mixed chaotic and quasiperiodic motions are used 
(dotted line (green stars): $C+QP$). These results indicate that for small coupling strength the Hamiltonian structure
still exists, but for larger values of coupling 
the  collective  dynamics becomes quasiperiodic; in this sense coupling induces simplicity in such coupled systems.

\subsection{Time delay coupling}
\begin{figure}
\includegraphics [scale=0.4]{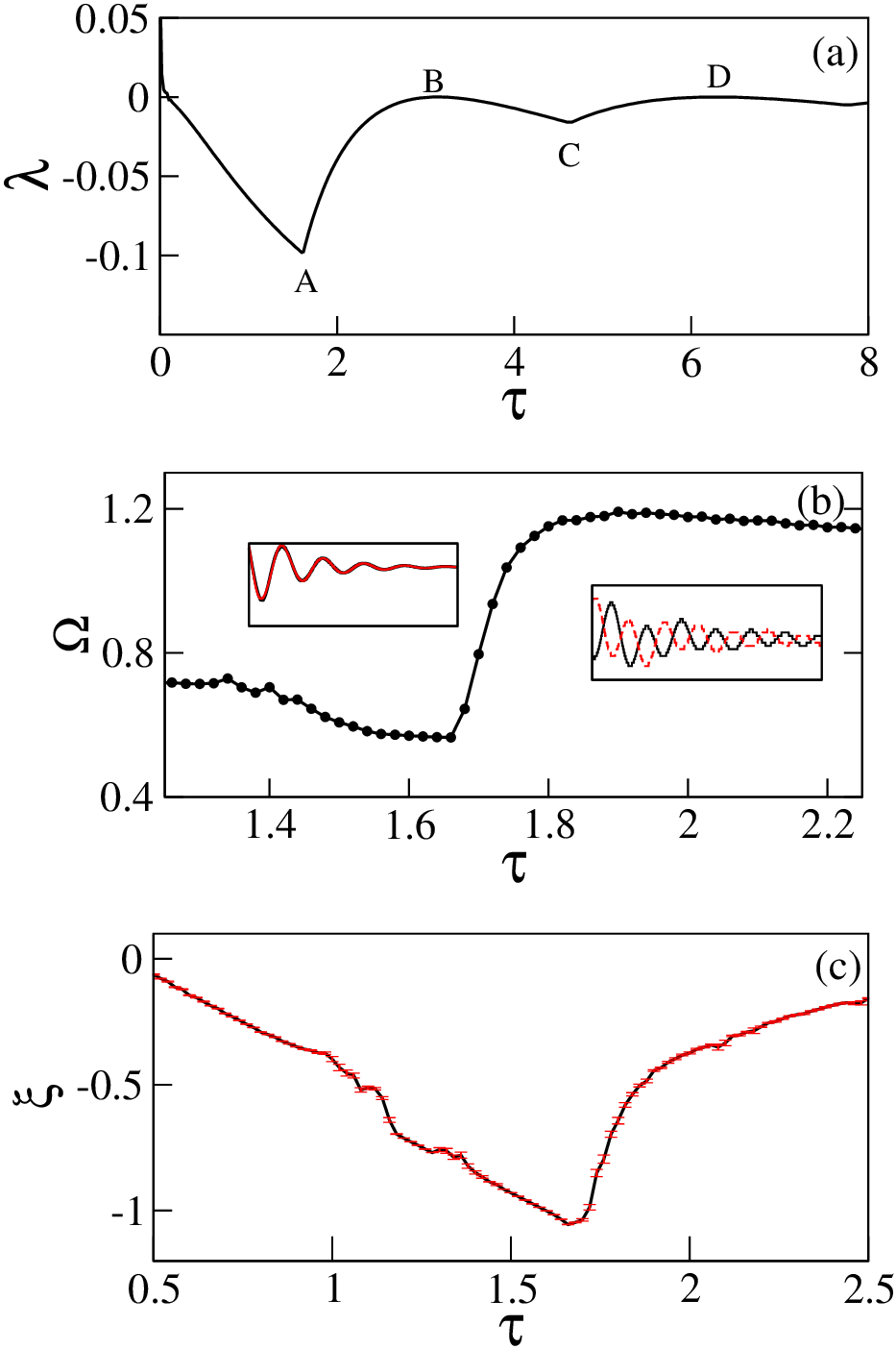}
\caption{(Color online)
(a) Largest Lyapunov exponent as a function of the time--delay, $\tau$, and
(b) the oscillation frequency, for coupling strength $\ep$ = 1. The inset
shows trajectories $x_1(t)$ and $x_2(t)$  for $\tau$ =1.4 and 1.8
respectively.  (c) The decay constant $\xi_1$ as a function of the delay
$\tau$.
}
\label{fig:hh1}
\end{figure}

When the two systems are coupled in presence of delay, $\tau$ (initial conditions taken from either the
regular or irregular motion) 
the largest Lyapunov exponent quickly becomes negative. This is shown as a
function of the delay  in Fig. \ref{fig:hh1}(a): almost as soon as the
delay is switched on the oscillators are driven to AD state.
The phenomenology of this higher dimensional system is very similar to that
of the coupled harmonic oscillators: the largest LE has the same
characteristic shape---there is a change in the slope at point A, where it
is clear that the phase-flip transition  occurs.  Shown in  Fig.
\ref{fig:hh1}(b) is the common frequency of oscillations as a function of
delay which changes discontinuously at $\tau \sim$1.65.
Since the phase of the oscillators is not clearly defined in such systems,
we infer the phase relation from the time--series (inset figures) of the
two systems  in the neighborhood of the point
of discontinuity. The phase difference changes from 0 to $\pi$ along with
the frequency jump as in the simpler 1--dimensional harmonic system.

Here also energy decreases exponentially in this AD region. We define the
energy of individual systems in the usual manner \cite{tabor}
\begin{eqnarray}
 E_j=
\frac{\dot{x_j}^2+\dot{y_j}^2}{2}+\frac{{x_j}^2+{y_j}^2}{2}+x_j^2y_j-\frac{y_j^3}{3},
\label{eq:ehh}
\end{eqnarray}
and quantify the energy dissipation as in Eq. (\ref{eq:en}) by a quantifier
$\xi_j$. The variation of this quantity with $\tau$ can be seen in Fig.
\ref{fig:hh1}(c). It confirms that energy dissipation is faster before 
phase flip transition and slower thereafter.

\begin{figure}
\vskip 0.8cm
\includegraphics [scale=0.4]{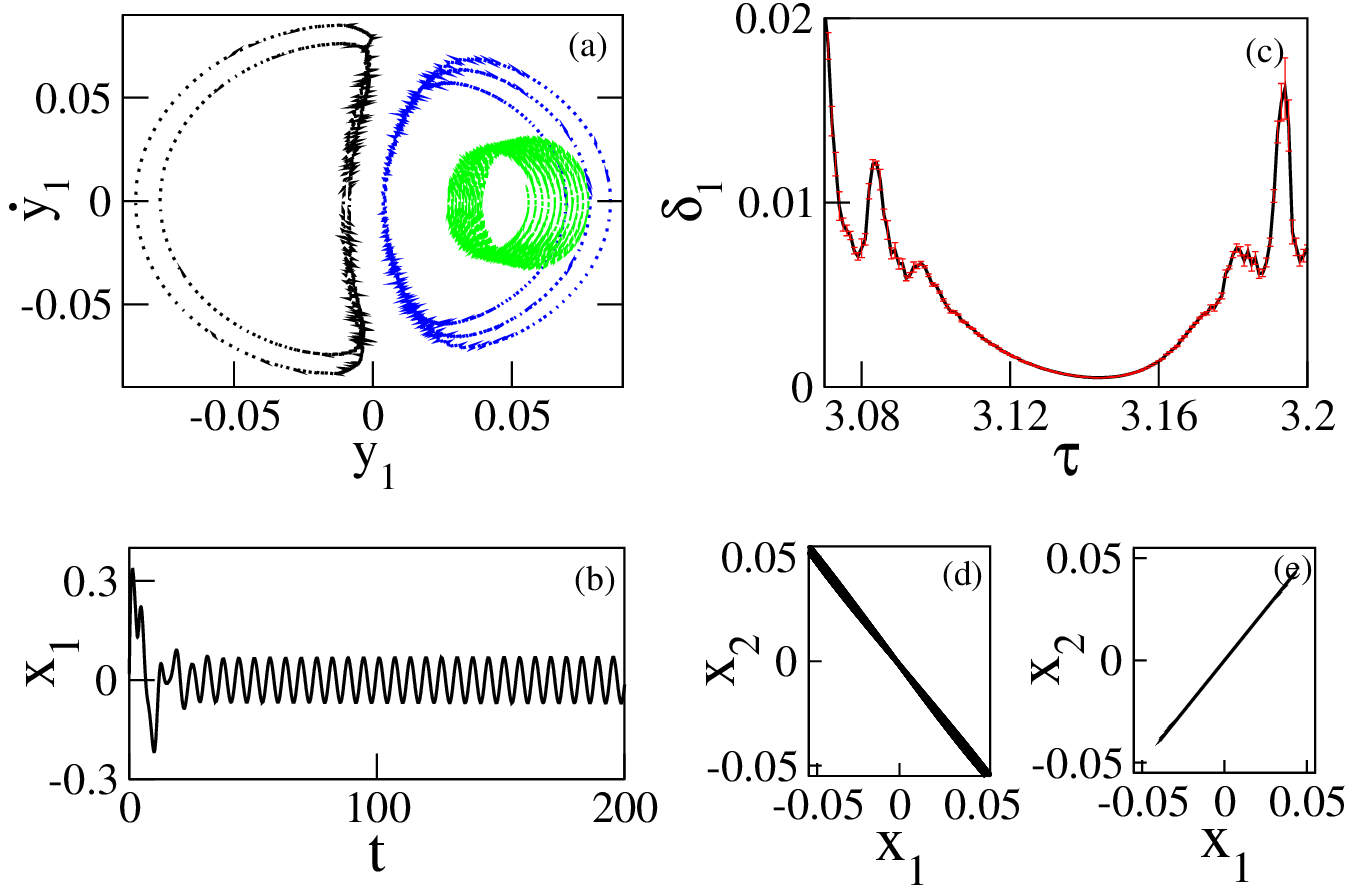}
\caption{ (Color online) (a) Poincar\'e section at $\tau=3.14$ for three
different initial conditions;
(b) time series of $x_1$; (c) variation of $\delta_1$ as a function of
delay.
The error bars are computed from a large number of initial conditions.  (d)
The relative phase in $x_{1}-x_{2}$ plane to show anti-phase motion at
$\tau=\pi$  and (e)
in-phase motion at $2\pi$.}
\label{fig:hh2}
\end{figure}

When the largest Lyapunov exponents approach zero (at points marked B and D in
Fig.~\ref{fig:hh1}(a)) the motion is oscillatory, decaying very slowly to
the fixed point.  Critical delays are at half the natural time period of oscillation,
and since the natural frequencies of the oscillators is equal to `1', the time period
is 2$\pi$. The Poincar\'e sections of representative trajectories at
$\tau=3.14$ are shown in Fig.~\ref{fig:hh2}(a) which are quasiperiodic (cf. Section IIA).
In the vicinity of this
point, the rate of decay can be computed numerically as discussed earlier,
and the quantifier $\delta$, Eq.~(\ref{eq:dis}) defined to locate the
oscillatory state also has a minimum for at $\tau \approx 3.14$. At
successive critical points ($\tau=3.14$ and $\tau=6.28$) the quasiperiodic
motions alternate in the nature of the phase synchrony; see
Fig.~\ref{fig:hh2}(d,e). Unlike the case of harmonic oscillators, though,
the coupling term does not quite vanish, so the emergence of oscillations is not
as pronounced in this case.

\section{Summary and Discussion }

In the present paper we have explored the effects of incorporating
time--delay in the coupling between Hamiltonian systems. The coupled system
effectively becomes dissipative, and in the absence of other attractors,
the total system displays amplitude death. However the coupling can effectively
vanish for specific values of the time--delay, at which delay the the system naturally 
appears to be conservative. We find that this behaviour differs from that of delay--coupled 
non--Hamiltonian systems, where AD islands are separated by finite ranges of delay values 
where the coupling function need  not vanish. 

Orbits near these critical delay values reflect
both dissipative  and conservative behavior: different initial conditions
give rise to different (seemingly) invariant curves which are very long
lived transients.
Energy dissipation in the AD regime is found to be associated with the phase--flip
 transition, and the damping is faster prior to the transition, and slower after it.

The finite velocity with which signals are transmitted gives rise to
intrinsic delays in the coupling, and as such this is germane in both
dissipative as well as conservative systems. Nevertheless, the effect of
delay has been explored to a limited extent in conservative systems. 
The results presented here may have more general applicability in coupled
systems with other conservation laws such as in ecological 
contexts \cite{eco}.

{\bf Acknowledgment:} GS gratefully acknowledges the support of the CSIR, India.  AP
and RR acknowledge
the  research support of the Department of Science and Technology (DST). AP also
acknowledges DST for DU-DST-Purse grant for financial support. This
work was started when AP was a sabbatical visitor at the MPI-PKS in
Dresden, Germany, and he acknowledges their kind hospitality.


\begin{thebibliography}{99}

\bibitem{sync}
L. M. Pecora and T. L. Carroll, \prl{64}{821}{1990}; S. Pikovsky, M. G. Rosenblum, and J. Kurths, 
\textit{Synchronization: A Universal Concept in Nonlinear Sciences} (Cambridge University Press,
Cambridge,  2001).

\bibitem{yam} Y. Yamaguchi and H. Shimizu, \phd{11}{212}{1984}; 
\bibitem{eli} K. Bar-Eli, \phd{14}{242}{1985}.
\bibitem{ins} D. G. Aronson, \etl, \phd{41}{403}{1990};
P. C. Mathews and S. H. Strogatz, \prl{65}{1701}{1990}; 
G. B. Ermentrout, \phd{ 41}{219}{1990}; 
\bibitem{resmi}V. Resmi, G. Ambika and R. E. Amritkar, \pre{84}{046212}{2011};
V. Resmi, G. Ambika, R. E. Amritkar and G. Rangarajan, \pre{85}{046211}{2012}. 
\bibitem{rep} G. Saxena, A. Prasad, and R. Ramaswamy, Phys. Reports {\bf 521},205 {(2012)}.

\bibitem{pf} A. Prasad, \pre{72}{056204}{2005}.
\bibitem{pfb}
A. Prasad, J. Kurths, S. K. Dana and R. Ramaswamy, \pre{74}{035204}{2006}; 
A. Prasad, S. K. Dana, R. Karnataka, J. Kurths, B. Blasius and R. Ramaswamy, Chaos {\bf 18}, {23111}, {(2008)}; 
C. Masoller, M. Torrent and J. O. Garcia, \ptr{367}{3255}{2009}; 
Y. Chen, J. Xiao, W. Liu, L. Li and Y. Yang, \pre{80}{046206}{2009}; 
J. M. Cruz, J. Escalona, P. Parmananda, R. Karnatak, A. Prasad, and R.
Ramaswamy, Phys. Rev. E. {\bf 81}, 046213 (2010); 
B. M. Adhikari, A. Prasad, M. Dhamala, Chaos {\bf 21}, 023116 {(2011)}; 
\bibitem{nirmal}
R. Karnatak, P. Nirmal, A. Prasad and R. Ramaswamy, Phys. Rev. E. {\bf 82}, 046219 (2010); 
P. Nirmal, R. Karnatak, A. Prasad, J. Kurths and R. Ramaswamy, \pre{85}{046204}{2012}.
\bibitem{hys} S. Kim, S. H. Park and C. S. Ryu, \prl{79}{2911}{1997}; 
A. Prasad, L. D. Iasemidid, S. Sabesan and K. Tsakalis, Pramana {\bf 64}, 513 
{(2005)}.

\bibitem{rid} J. C. Sommerer and E. Ott, Nature {\bf 365}, 138 {(1993)}.

\bibitem{gen}  K. Kaneko, {\sl Theory and Applications of Coupled Map
Lattices} (John Wiley and Sons, New York, 1993).

\bibitem{wagner} H. G. Schuster and P. Wagner, \ptp{81}{939}{1989}.
\bibitem{konishi} K. Konishi, M. Ishii and H. Kokame, \pre{54}{3455}{1996}; 
K. Konishi, H. Kokame and K. Hirato, \pre{62}{384}{2000}; 
K. Konishi, \pre{67}{017201}{2003}; 
K. Konishi, K. Senda and H. Kokame, \pre{78}{056216}{2008}; 
K. Konishi, H. Kokame and N. Hara, \pre{81}{016201}{2010}; 
L. B. Le, K. Konishi and N. Hara, Nonlinear Dyn. {\bf 67}, 1407,{(2012)}.
\bibitem{reddyall} D. V. Ramana Reddy, \etl, \prl{80}{5109}{1998}; 
S. H. Strogatz, Nature (London) {\bf 394}, 316 (1998); 
F. M. Atay, \prl{91}{094101}{2003};  
F. M. Atay, J. Jost, \prl{92}{144101}{2004}; 
F. M. Atay, \phd{183}{1}{2003}; 
G. Saxena, A. Prasad, and R.Ramaswamy, Phys. Rev. E. {\bf 82}, 017201 {(2010)}.
\bibitem{zou}W. Zou, C. Yao and M. Zhan, \pre{82}{056203}{2010}; 
W. Zou, J. Lu, Y. Tang, C. Zhang and J. Kurths, \pre{84}{066208}{2011}; 
W. Zou, Y. Tang, L. Li and J. Kurths, \pre{85}{046206}{2012}; 
W. Zou, D. V. Senthilkumar, Y. Tang and J. Kurths, \pre{86}{036210}{2012}.

\bibitem{ravi}V. Ravichandran, V. Chinnathambi and S. Rajshekar, Pramana {\bf 78}, 2911 {(2012)}.
\bibitem{senthil} M. Lakshamanan and D. V. Senthilkumar, {\sl Dynamics of Nonlinear Timedelay systems} 
(Springer, India, 2011).

\bibitem{farmer} J. D. Farmer, \phd{4}{366}{1982}.

\bibitem{tabor} M. Tabor, {\sl Chaos and Integrability in Nonlinear
Dynamics: An Introduction} (John Wiley \& Sons,  New York, 1989).
\bibitem{arnold} V. I. Arnold, {\it Mathematical Methods of Classical Mechanics} (Springer, New York, 1978).
\bibitem{zanette} D. H. Zanette and A. S. Mikhailov, \pla{235}{135}{1997};
A. Hampton and D. H. Zanette, \prl{83}{2197}{1999}.

\bibitem{wang} Z. Wang and H. Hu, Proceedings of International Design
Engineering Technical
Conferences \& Computers and Information in Engineering Conference  {(2005)}.

\bibitem{mis} Similar results are found for mismatched oscillators. 

\bibitem{num}
The flow is integrated using the Runge-Kutta $4^{th}$ order scheme
with integration step $\Delta t = \tau/N$ where $N = 300$ is fixed.
As we increase the value $N$ (checked upto  $N=1000$) the decrease in
amplitude in Fig. 3(b) becomes slower. Lyapunov exponents are calculated using the 
method as given in Ref. \cite{farmer}.
\bibitem{fqcal} Phase and  frequency are  numerically calculated as per Ref. \cite{pf}. 


\bibitem{trans} Understanding the transient dynamics is essential in
describing AD since this
state is asymptotically featureless.  Transients can be  significant in
applications that are restricted to finite times, as for example in ecology \cite{tra}.

\bibitem{tra} A. Hatings, Eco. Lett. {\bf 215}, 4 {(2001)}; O. Ovaskainen, 
I. Hanski Theor. Popul. Biol. {\bf 61}, 285 {(2002)}; 

\bibitem{eco} Y. Nutku, \pla{145}{27}{1990}.


\end{thebibliography}
\end{document}